\documentclass[cameraready]{Interspeech}

\title{Towards Paradigm-General Suicide Risk Detection via Speech LLM}

\author[affiliation={2}, orcid=0009-0006-7472-8258, equalcontribution]{Jialun}{Li}
\author[affiliation={2}, orcid=0009-0003-1743-7428, equalcontribution]{Weitao}{Jiang}
\author[affiliation={1,2}, orcid=0009-0009-0485-6141]{Ziyun}{Cui}
\author[affiliation={3}, orcid=0000-0002-5934-7346]{Yinan}{Duan}
\author[affiliation={3}, orcid=0000-0002-9999-8725]{Diyang}{Qu}
\author[affiliation={1,2}, orcid=0000-0002-7730-5131]{~\\Chao}{Zhang}
\author[affiliation={3}, orcid=0000-0002-2145-8630]{Runsen}{Chen}
\author[affiliation={3}, orcid=0000-0003-3506-0110, correspondingauthor]{Chang}{Lei}
\author[affiliation={1}, orcid=0000-0001-8116-7715, correspondingauthor]{Wen}{Wu}

\address{
    $^1$ Shanghai Artificial Intelligence Laboratory \\
    $^2$ Department of Electronic Engineering, Tsinghua University \\
    $^3$ Vanke School of Public Health, Tsinghua University
}

\email{leic22@mails.tsinghua.edu.cn, wuwen@pjlab.org.cn}


\usepackage{amsmath,amssymb}
\usepackage{graphicx}
\usepackage{multirow}
\usepackage{booktabs}
\usepackage{url}
\usepackage{xcolor}

\usepackage{cite}

\begin{document}

\maketitle

\begin{abstract}

    Suicide risk among adolescents remains a critical public health concern, and speech provides a non-invasive and scalable approach for its detection. 
    Speech-based suicide risk assessment commonly relies on carefully designed speech elicitation paradigms (\textit{e.g.,} verbal fluency, reading, or question answering) to probe cognitive and affective states. Existing approaches, however, typically focus on one single paradigm at a time. This paper, for the first time, investigates cross-paradigm approaches that unify diverse speech elicitation paradigms within a single model. Specifically, we use a speech LLM as backbone with a mixture of DoRA experts (MoDE) to capture complementary cues across assessments dynamically, tested on 1,223 participants across ten speech elicitation paradigms. Results show that MoDE outperforms both paradigm-specific and conventional joint-learning models. Moreover, it can generalise to unseen paradigms and provide better confidence calibration.
    
\end{abstract}

\section{Introduction}
\label{sec:intro}

Suicide remains a leading cause of death among adolescents, making timely risk identification a public health imperative~\cite{suicide111}. Detection of suicide risk is challenging not only because it lacks a uniform clinical profile but also because it often depends on patients' willingness and ability to disclose subjective experiences~\cite{CUMMINS201510}. Speech offers a cost-efficient, remote, and non-invasive window into mental states. As individuals approach a suicide crisis, measurable changes occur in acoustic dimensions, such as jitter, shimmer, fundamental frequency, and disfluency~\cite{HOMAN2022102161}, as well as lexical markers~\cite{belouali2021acoustic}. 

Previous studies have shown the potential of speech as a modality for suicide risk detection~\cite{cui24_interspeech,wu25_interspeech}. Participants are typically guided through several structured interaction protocols designed to elicit speech under specific cognitive or affective conditions, such as self-description and passage reading, which we refer to as speech elicitation paradigms (SEP) in this paper. A model then performs detection by taking those speech recordings as input. Existing approaches often focus on a single SEP at a time~\cite{cui24_interspeech}, which requires separate models for each SEP. Empirical evidence suggests that different SEP elicit complementary acoustic and semantic cues~\cite{belouali2021acoustic,lin2025machine}, motivating the development of cross-paradigm frameworks capable of integrating these complementary cues~\cite{DBLP:journals/corr/abs-1708-03920,parsapoor2023performance}. 

The 1st SpeechWellness Challenge~\cite{wu25_interspeech} incorporates three SEPs for automatic suicide risk detection. Results from the challenge demonstrated that ensembling information from multiple SEPs improves both predictive accuracy and robustness. Furthermore, speech LLMs have shown advanced performance on a variety of tasks such as end-to-end translation, speech recognition and emotion recognition~\cite{pu2025empowering,tang2023salmonn,rubenstein2023audiopalm}. The strong generalisability of such large models provides new opportunities for cross-paradigm suicide risk detection.

This paper investigates paradigm-general suicide risk detection using speech LLMs. A novel structure, named Mixture of DoRA Experts (MoDE), is adopted to unify diverse SEPs into a single model. Mixture of experts (MoE) have emerged as a powerful strategy for LLMs~\cite{masoudnia2014mixture,li2024locmoe,artetxe2022efficient,xie2024moe,xue2024openmoe}, which decompose the problem into specialised expert subnetworks. Compared to traditional MoE, where each expert is a full subnetwork, this paper employs MoDE which uses weight-decomposed low-rank adaptation (DoRA)~\cite{liu2024doraweightdecomposedlowrankadaptation} as experts for a more lightweight and efficient design.
The proposed method was tested on 10 SEPs for automatic suicide risk detection. Results show that MoDE successfully integrates 10 paradigms into a single model by dynamically capturing paradigm-specific and cross-paradigm shared information across experts. MoDE outperforms paradigm-specialised models as well as conventional joint tuning methods. More importantly, the trained paradigm-general model demonstrates generalisation capability to unseen speech elicitation paradigms, successfully transferring knowledge learned from multiple paradigms to novel assessment settings not observed during training. Furthermore, confidence analysis and reject option experiments show that cross-paradigm learning also improves confidence calibration, which is important for medical-related tasks. To the best of our knowledge, this is the first work that investigates suicide risk detection using speech LLM and is also the first to unify multiple speech suicide risk elicitation paradigms within a single model.

\section{Methods}
\label{sec:method}

\subsection{Datasets and Speech Elicitation Paradigms}

The dataset used in this study~\cite{cui24_interspeech,wu25_interspeech} consists of voice recordings collected from 1,223 Chinese adolescents (aged 10--18). Data were obtained under controlled experimental conditions and paired with standardised suicide risk assessments administered via the Mini International Neuropsychiatric Interview for Children and Adolescents (MINI-KID) suicidality module~\cite{cite104088JCP09m05305whi}, a widely validated diagnostic interview tool. Among the participants, 53.4\% of them were identified as in suicide risk. Each participant completed ten structured SEPs designed to elicit a broad range of linguistic, cognitive, and emotional features. Details of the ten SEPs are described in Table~\ref{tab:task}. 

This dataset provides a clinically annotated speech resource for investigating adolescent suicide risk. The data collection procedures have been approved by the Medical Ethics Committee of the Tsinghua University Technology Ethics Committee.

\begin{table}[tbp]
\centering
\caption{Description of the ten SEPs.}
\label{tab:task}
\vspace{-1.5ex}
\resizebox{0.85\linewidth}{!}{
\begin{tabular}{cl}
\toprule
\textbf{SEP} & \textbf{Description} \\
\midrule
\textbf{01} & Animal Fluency Test~\cite{whiteside2016verbal} \\
\textbf{02} & Open-ended question: self-description \\
\textbf{03} & Open-ended question: happy memory \\
\textbf{04} & Open-ended question: manage distress \\
\textbf{05} & Passage reading~\cite{baird2022blowing} \\
\textbf{06} & Word reading~\cite{xu2008construction,zhang2010study} \\
\textbf{07} & Describe face with negative emotion~\cite{conley2018racially} \\
\textbf{08} & Describe face with positive emotion~\cite{conley2018racially} \\
\textbf{09} & Describe face with neutral emotion~\cite{conley2018racially} \\
\textbf{10} & Generate possible uses for an empty box~\cite{runco2012divergent} \\
\bottomrule
\end{tabular}
}
\end{table}

\subsection{Speech LLM with DoRA}

\begin{figure}[t]
    \centering
    \includegraphics[width=\linewidth]{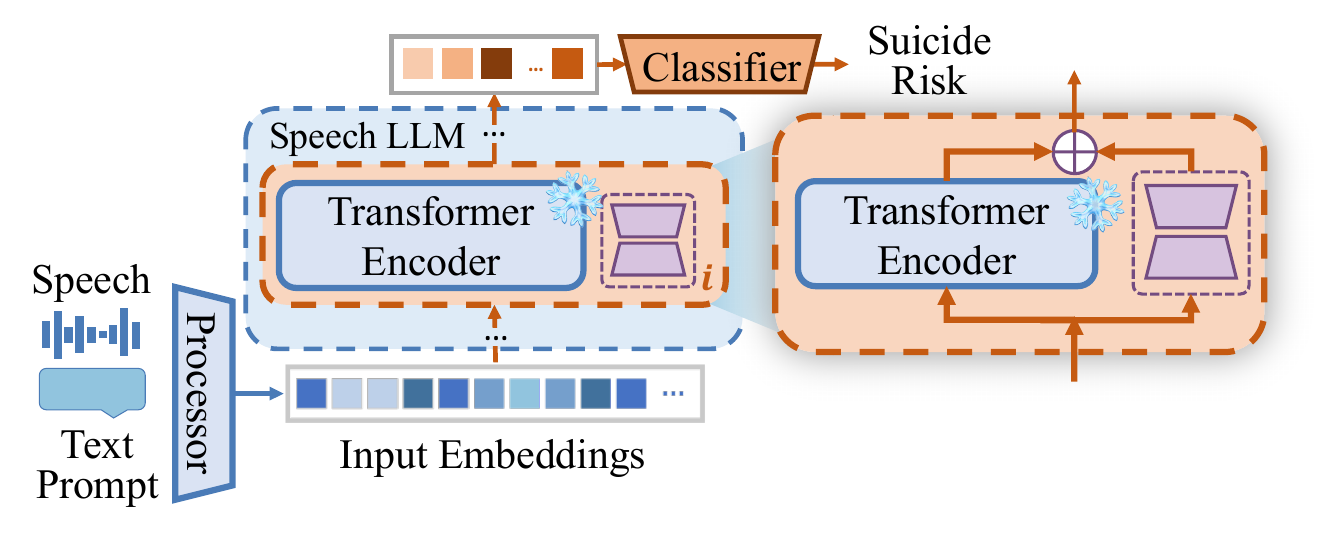}
    \vspace{-4ex}
    \caption{Architecture of the Baseline model}
    \label{fig:arch-baseline}
    \vspace{-2ex}
\end{figure}

As shown in Fig.~\ref{fig:arch-baseline}, a speech LLM is employed as the backbone model to process speech recordings and generate embeddings for downstream suicide risk detection. Weight-decomposed low rank adaptation (DoRA)~\cite{liu2024doraweightdecomposedlowrankadaptation} is adopted for parameter-efficient finetuning. DoRA decomposes the pretrained weight into its magnitude and directional components, and applies low-rank updates only to the direction while fine-tuning both components. It can be formulated as
\begin{equation}
    W'=m \frac{W_0+BA}{||W_0+BA||_c},
\end{equation}
where $W_0$ denotes the pretrained weight, $B$ and $A$ are two low-rank matrices, $m=||W_0||_c$ is a trainable scalar, and the product $BA$ represent the update of direction. DoRA has been shown to outperform LoRA~\cite{hu2021loralowrankadaptationlarge} on various speech- and language-related tasks~\cite{liu2024doraweightdecomposedlowrankadaptation}.

\begin{figure}[t]
    \centering
    \includegraphics[width=\linewidth]{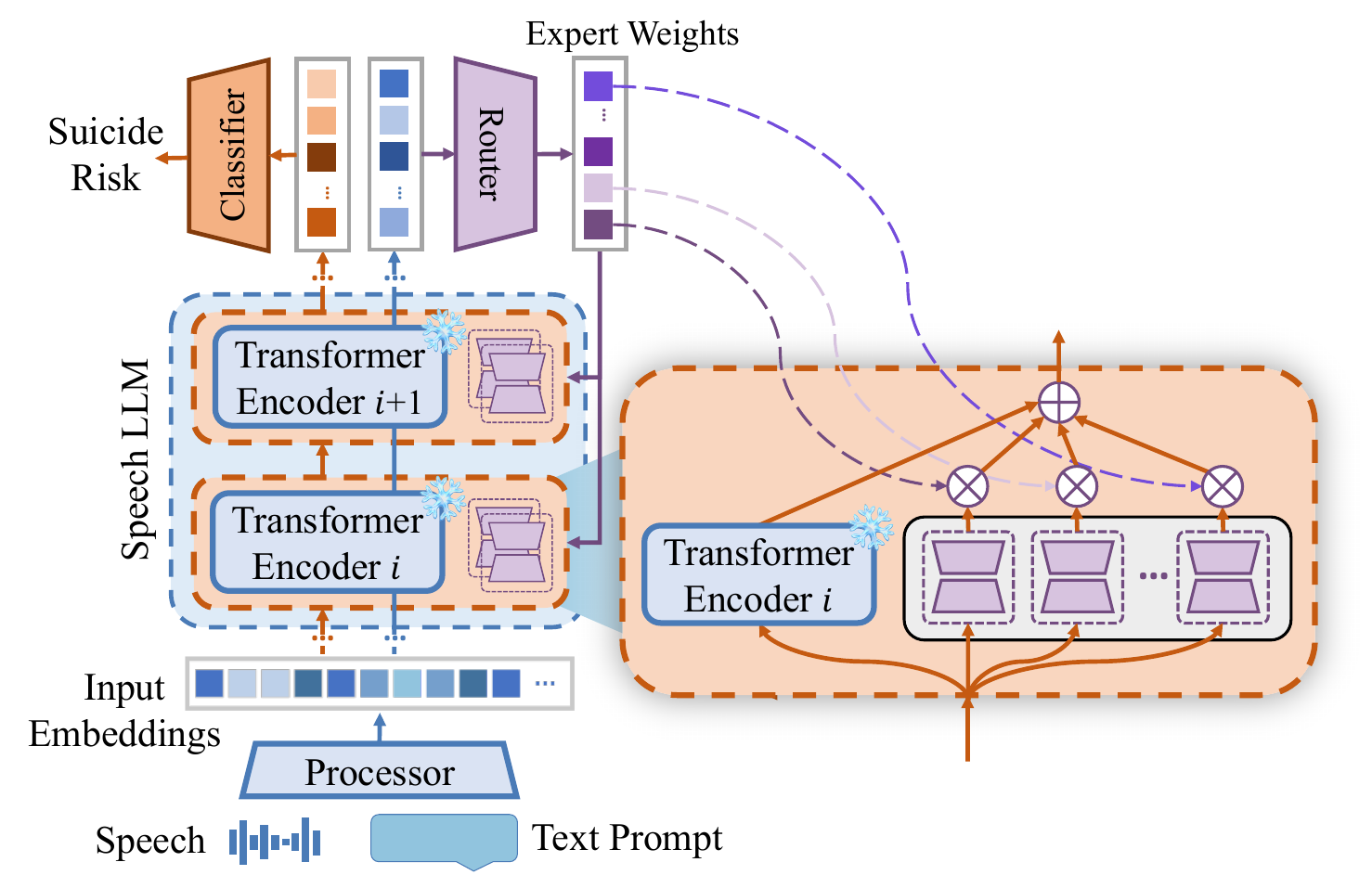}
    \vspace{-3ex}
    \caption{Architecture of the MoDE model}
    \label{fig:arch-mode}
    \vspace{-1ex}
\end{figure}

\subsection{Mixture of DoRA Experts}

We propose the mixture of DoRA experts (MoDE) for cross-paradigm suicide risk detection. As shown in Fig.~\ref{fig:arch-mode}, MoDE replaces the single adapter with multiple specialised experts and a dynamic routing mechanism. The last hidden states of the Transformer block are fed into a fully connected router to generate expert weights $\mathbf{w}=\text{softmax}(\mathbf{R(h)})$, where $\mathbf{h}$ denote the last hidden states, $\mathbf R$ denotes the router, and $\mathbf w=\begin{bmatrix}w_1\,,w_2\,,\cdots\,,w_E\end{bmatrix}^T$ serves as expert weights for $E$ experts.

In MoDE, the weight matrices are updated through a weighted combination of DoRA experts:
\begin{equation}
    W' = m \frac{W_0 + \sum_{i=1}^{E} w_i (B_i A_i)}{||W_0 + \sum_{i=1}^{E} w_i (B_i A_i)||_c}
\end{equation}
where $w_i$ are the experts weights, and $A_i,B_i$ denote the DoRA-specific weights for $i^\text{th}$ expert.

To improve training stability, the router first computes expert weights using the backbone without DoRA modules, after which the DoRA experts are activated with these fixed weights to produce the final prediction. This decoupled design prevents the router from being affected by adaptation dynamics of DoRA experts, leading to more stable and reliable expert selection.

\subsection{Load Balancing and Temperature Scaling}

Load balancing is further introduced to balance expert utilisation and prevent expert collapse, which adds a KL divergence between expert weights $\mathbf{w}$ and expected weight contribution as the auxiliary loss. Specifically, a uniform distribution $\mathbf{u} = [1/E, 1/E, \ldots, 1/E]^T$ is used as the constraint weight, where $E$ is the number of experts. The final loss is formulated as:
$\mathcal{L} = \mathcal{L_\text{CE}} + \lambda_\text{LB} D_\text{KL}(\mathbf{w} || \mathbf{u})$,
where the $\mathcal{L_\text{CE}}$ is cross-entropy loss for classification, and the $\lambda_\text{LB}$ denotes the coefficient of the auxiliary load balancing loss.

Temperature scaling is a post-processing calibration method~\cite{pmlr-v70-guo17a} which is applied to the softmax normalisation of the router  weights. By adding temperature scaling, the model can control the sparseness of expert selection and balance the expert specialisation and diversity.

\section{Experimental Setup}
\label{sec:exps}

\subsection{Backbone Model and Baselines}

Qwen2.5-Omni-7B~\cite{xu2025qwen2}, a multimodal large language model capable of processing both speech and text inputs, was used as the backbone model. To evaluate the effectiveness of speech LLMs on suicide risk detection, we compared Qwen-Omni with Whisper-Large-v3~\cite{radford2023robust}, a widely used speech foundation model. Three MLP layers were added on top of the Whisper model for classification. The task is formulated as a binary detection task. Instruction tuning was applied for the Qwen-Omni model with prompt template shown as follows: \emph{``The speaker responded to the following question: \{Task Description\}. Based on the provided audio, assess the suicide risk of the speaker (Yes, No).''}

The proposed MoDE method was compared to the following baselines: (i) \emph{Separate tuning}: A separate model was trained for each single paradigm using data from the correspondent paradigm; (ii) \emph{Joint tuning}: The model was trained on data from all 10 SEPs. Since the dataset is roughly balanced, performance was evaluated using accuracy.

\subsection{Implementation Details}

The dataset was split into train/dev/test sets at a ratio of 8:1:1. To ensure comparability of results across all ten paradigms, the test set was subsequently refined to include only participants who had completed every paradigm. The system was implemented using PyTorch, with the AdamW optimiser and the cosine learning rate scheduler. The model was trained for 4 epochs with a batch size of 64. MoDE uses 10 DoRA experts with rank of 32 and alpha of 64. The model achieving the best validation performance was used for testing evaluation. All experiments were conducted with three random seeds. Both mean and standard error are reported.

\section{Results and Discussion}
\label{sec:result}

\subsection{Using Speech LLM for Suicide Risk Detection}

\begin{figure}[t!]
    \centering
    \includegraphics[width=\linewidth]{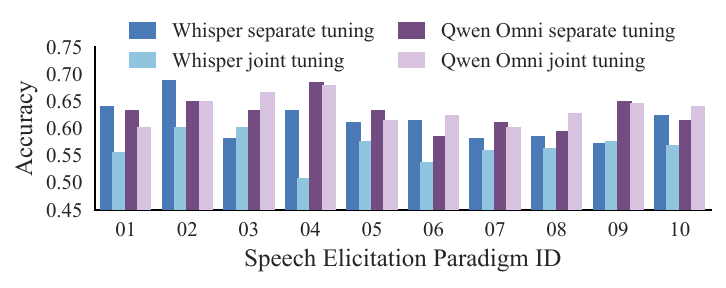}
    \vspace{-4ex}
    \caption{Performance of baseline models on 10 SEPs.}
    \label{fig:baseline}
\end{figure}

We first compared Qwen-Omni with Whisper on both \emph{separate tuning} and \emph{joint tuning} settings to evaluate the effectiveness of speech LLM on suicide risk detection tasks. Results are shown in Fig.~\ref{fig:baseline}. It can be observed that, under the separate tuning setting, Qwen-Omni does not necessarily outperform Whisper. However, when joint tuning using all SEPs, Qwen-Omni yields higher accuracy than Whisper on 8 out of 10 SEPs, demonstrating superior multitasking capability. In addition, the results indicate that using a joint tuning strategy that simply mixes data from difference paradigms can hardly yield consistent improvement across paradigms.

\subsection{Performance of MoDE for Cross-Paradigm Tuning}

The results of the proposed MoDE method for joint tuning are shown in Table~\ref{tab:results}. It can be seen that introducing MoDE to joint tuning boosts the performance on most paradigms and leads to a 4.5\% relative improvement in average accuracy compared with separate tuning. The results demonstrate that MoDE not only enables a single model to perform ten paradigms efficiently, but also achieves superior performance compared with paradigm-specific models.

\begin{table}[t]
\centering
\caption{Accuracy of 10 speech elicitation paradigms. Experiments conducted with three different seeds. Average and standard error reported. Best results shown in bold.}
\label{tab:results}
\vspace{-1ex}
\resizebox{0.95\linewidth}{!}{
\begin{tabular}{@{}cccc@{}}
\toprule
\textbf{SEP} & \textbf{Separate Tuning} & \textbf{Joint Tuning} & \textbf{Joint with MoDE} \\ \midrule
\textbf{01} & 0.632$\pm$0.016 & 0.602$\pm$0.023 & \textbf{0.636$\pm$0.013} \\
\textbf{02} & 0.649$\pm$0.007 & 0.649$\pm$0.000 & \textbf{0.667$\pm$0.019} \\
\textbf{03} & 0.632$\pm$0.016 & 0.667$\pm$0.011 & \textbf{0.671$\pm$0.022} \\
\textbf{04} & 0.684$\pm$0.009 & 0.680$\pm$0.009 & \textbf{0.758$\pm$0.009} \\
\textbf{05} & 0.632$\pm$0.011 & 0.615$\pm$0.011 & \textbf{0.658$\pm$0.019} \\
\textbf{06} & 0.584$\pm$0.007 & 0.623$\pm$0.013 & \textbf{0.632$\pm$0.009} \\
\textbf{07} & 0.610$\pm$0.020 & 0.602$\pm$0.016 & \textbf{0.632$\pm$0.011} \\
\textbf{08} & 0.593$\pm$0.009 & 0.628$\pm$0.016 & \textbf{0.675$\pm$0.013} \\
\textbf{09} & \textbf{0.649$\pm$0.026} & 0.645$\pm$0.009 & 0.636$\pm$0.026 \\
\textbf{10} & 0.615$\pm$0.022 & \textbf{0.641$\pm$0.024} & 0.593$\pm$0.011 \\ \midrule
\textbf{Avg.} & 0.628$\pm$0.006 & 0.635$\pm$0.005 & \textbf{0.656$\pm$0.003} \\ \bottomrule
\end{tabular}
}
\end{table}

\subsection{Ablation Study of MoDE modules}
\label{sec:ablation}

This section examines the effect of each MoDE module. As shown in Table~\ref{tab:ablation}, removing temperature scaling results in a 2.4\% relative performance drop. Moreover, it is worth noting that without load balancing, MoDE collapses to a single DoRA expert, leading to further degradation in performance.

\begin{table}[tb]
\centering
\caption{Ablation study of MoDE models. Average accuracy over 10 speech elicitation paradigms reported.}
\label{tab:ablation}
\vspace{-1ex}
\setlength{\tabcolsep}{4pt} 
\resizebox{\linewidth}{!}{
\begin{tabular}{lccc}
\toprule
\textbf{Model} & \textbf{MoDE} & \textbf{-- Temp. Scaling} & \textbf{-- Load Balancing} \\
\midrule
\textbf{Avg. Acc.} & {0.656$\pm$0.003} & 0.640$\pm$0.006 & 0.625$\pm$0.010 \\
\bottomrule
\end{tabular}
}
\end{table}

\subsection{Varying the Number of Experts}

\begin{figure}[t]
\centering
\includegraphics[width=0.85\linewidth]{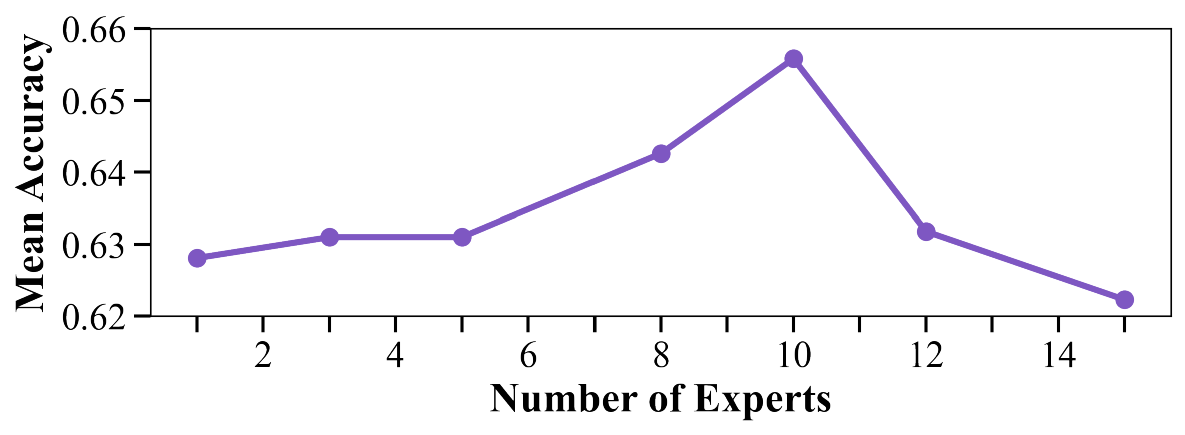}
\vspace{-1ex}
\caption{Performance of MoDE with different number of experts. Average accuracy over 10 SEPs reported.}
\label{fig:accuracy_vs_experts}
\end{figure}

This section investigates the effect of the number of experts. Fig.~\ref{fig:accuracy_vs_experts} reports the average accuracy across 10 SEPs under a range of numbers of experts. Performance initially improves as the number of experts increases but eventually declines beyond a certain point. This trend can possibly be attributed to two competing factors: on one hand, additional experts increase model capacity and enhance the ability to capture diverse paradigm categories; on the other hand, too many experts introduce routing noise and reduce training effectiveness, as each expert receives insufficient data.

\subsection{Manual Supervision on Expert Assignment}

In this section, we further analyse whether a manually defined prior is useful to guide expert specialisation. Specifically, we adopted a two-stage training pipeline. In Stage I, each paradigm was explicitly assigned to a designated expert, pre-training the experts to develop paradigm-specific capabilities. In Stage II, the router was unfrozen and all experts were jointly trained. Results show that this strategy yields an average accuracy of 0.640, lower than the proposed MoDE approach where routing is learned automatically without manual supervision. These results indicate that although manual supervision on expert assignment can enhance some underperforming paradigms through targeted specialisation, it underutilises potential inter-paradigm collaborations. The fixed paradigm-expert assignment in Stage I imposes a strong prior that limits the router's ability to dynamically capture cross-paradigm relationships. As a result, the model fails to achieve the additional gains seen in fully unsupervised routing, where the router can freely exploit latent synergies between paradigms.

\subsection{Paradigm-Expert Specialisation Analysis}

\begin{figure}[t]
    \centering
    \begin{minipage}[b]{0.48\linewidth}
        \centering
        \includegraphics[width=\linewidth]{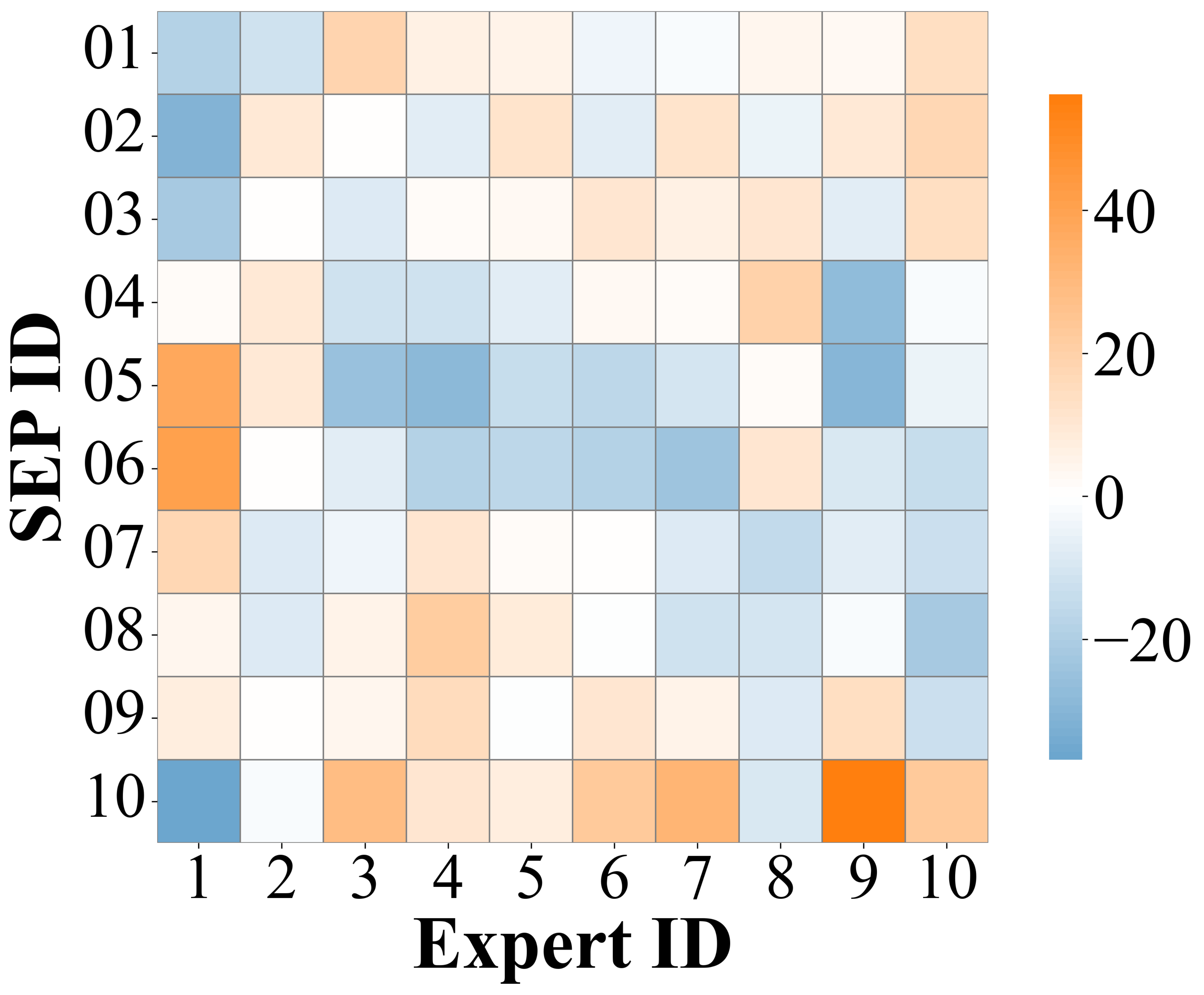}
        \caption{Activation heatmap of experts.}
        \label{fig:expert_activation_heatmap}
    \end{minipage}
    \hfill
    \begin{minipage}[b]{0.48\linewidth}
        \centering
        \includegraphics[width=0.98\linewidth]{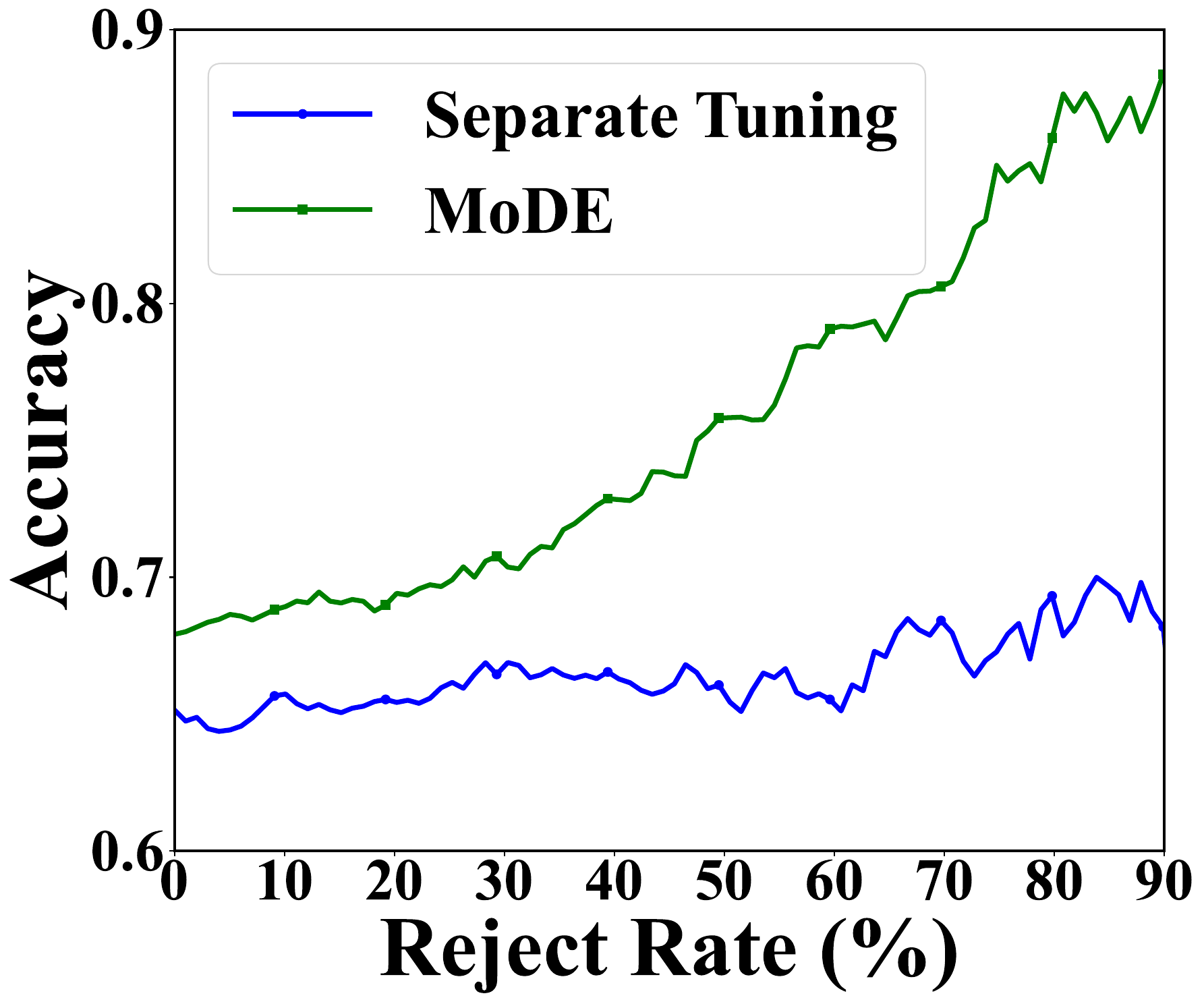}
        \caption{Reject option based on confidence scores.}
        \label{fig:rejection-curve}
    \end{minipage}
    \vspace{-3ex}
\end{figure}

We further examine the routing activation patterns to better understand the paradigm-expert correlation. Results are plotted in Fig.~\ref{fig:expert_activation_heatmap}. For reading tasks SEP05 and SEP06, where acoustic features dominate semantic meaning, we observe strong activation of expert E0 while consistent suppression of experts E2-4 and E8. Similarly, image description tasks SEP07-09 primarily engage expert E3 with reliable inhibition of experts E1, E7 and E9. Notably, SEP10 exhibits fundamentally distinct activation characteristics from all other tasks, engaging a unique combination of experts E2, E5, E6, E8 and E9 while strongly suppressing expert E0. This pattern aligns with its exceptional position in our agreement analysis and confirms its fundamentally different nature within the paradigm set. This analysis reveals that the proposed method learns meaningful paradigm–expert relationships without explicit supervision, demonstrating the effectiveness of the routing mechanism in automatically discovering optimal paradigm–expert assignments. This finding echoes the results from the previous section that automatic routing outperforms explicit expert assignment.

\begin{table}[t]
    \centering
    \caption{Zero-shot generalisation on unseen SEPs. MoDE is trained on nine SEPs and evaluated on an unseen one, compared to original speech LLM without cross-paradigm training.}
    \label{tab: zero-shot}
    \vspace{-1ex}
    \resizebox{0.92\linewidth}{!}{
    \begin{tabular}{ccc}
    \toprule
         \textbf{Unseen SEP} &  \textbf{Original backbone} &  \textbf{Backbone w/ MoDE}\\
         \midrule
         SEP05& 0.286  &  0.616 $\pm$ 0.013\\
         SEP07 & 0.156  &  0.593 $\pm$ 0.008 \\
         \bottomrule
    \end{tabular}
    }
    
\end{table}

\subsection{Generalisation to Unseen Paradigms}
To assess the generalisation capability of the proposed cross-paradigm training approach, we perform a leave-one-paradigm-out evaluation. MoDE is trained on nine SEPs and tested in a zero-shot manner on a held-out, unseen paradigm, with zero-shot inference of the original pretrained speech LLM used as a baseline to control for inherent backbone capability.
As shown in Table~\ref{tab: zero-shot}, the original backbone fails to achieve meaningful suicide risk detection on unseen paradigms. In contrast, MoDE exhibits strong cross-paradigm generalisation, attaining performance close to that of models separately tuned on the target paradigm (first column of Table~\ref{tab:results}). These results demonstrate that cross-paradigm learning enables the acquisition of paradigm-agnostic representations, supporting robust suicide risk detection under previously unseen assessment conditions.


\subsection{Cross-Paradigm Learning Improves Confidence Calibration}

This section evaluates the prediction reliability of the proposed method with respect to confidence calibration. Effective calibration ensures that a model's predicted probabilities accurately reflect the actual likelihood of correctness. We assess calibration performance using the following metrics: (i) expected calibration error (ECE)~\cite{naeini2015obtaining} and maximum calibration error (MCE)~\cite{siu1997improved} which quantify the deviation between confidence and accuracy; (ii) negative log-likelihood (NLL) and normalised cross-entropy (NCE) that evaluate probabilistic prediction quality; (iii) area under the ROC curve (AUROC) which measures separability between correct and incorrect predictions; (iv) area under the precision-recall curve (AUPRC) that assesses precision-recall trade-off across confidence thresholds. Softmax probability of the predicted class is used as confidence score. As shown in Table~\ref{tab:confidence}, MoDE achieves consistently better results across all metrics, indicating that it  not only improves detection accuracy but also enhances confidence calibration.

\begin{table}[t]
\centering
\caption{Confidence calibration performance of separate tuning and MoDE models.}
\label{tab:confidence}
\vspace{-1ex}
\setlength{\tabcolsep}{4pt} 
\resizebox{\linewidth}{!}{
\begin{tabular}{@{}lcccccc@{}}
\toprule
\textbf{Model} & \textbf{ECE$\downarrow$} & \textbf{MCE$\downarrow$} & \textbf{NLL$\downarrow$} & \textbf{NCE$\uparrow$} & \textbf{AUROC$\uparrow$} & \textbf{AUPRC$\uparrow$} \\ \midrule
\textbf{Separate} & 0.099 & 0.225 & 0.805 & -0.235 & 0.640 & 0.683 \\
\textbf{MoDE} & \textbf{0.061} & \textbf{0.089} & \textbf{0.645} & \textbf{0.013} & \textbf{0.686} & \textbf{0.706} \\ \bottomrule
\end{tabular}
}
\vspace{-2ex}
\end{table}

We further conduct a confidence-based rejection analysis, where predictions with confidence below a threshold are withheld and the accuracy on the retained subset is computed. As shown by the rejection curve in Fig.~\ref{fig:rejection-curve}, MoDE’s accuracy rises with higher rejection rates, reflecting that predicted confidence scores successfully reveal error-prone predictions. In contrast, the separate tuning model yields a relatively flat curve. The analysis shows that MoDE enhances the reliability for decision-making, which is critical for medical applications.

\section{Conclusions}

For automatic suicide risk detection from spontaneous speech, participants are typically asked to complete multiple speech elicitation paradigms, and detection models operate on the resulting recordings as input. This paper investigates unifying diverse speech elicitation paradigms within a single model for paradigm-general suicide risk detection. A novel architecture, MoDE, is adopted which employs a speech LLM as the backbone and incorporates a mixture of DoRA experts to dynamically capture both paradigm-specific and cross-paradigm shared information. Experiments on 1,223 participants across 10 speech elicitation paradigms demonstrate that the proposed method outperforms paradigm-specific models and conventional joint-tuning approaches. 
Furthermore, we also found that cross-paradigm training provides generalisability to unseen paradigms as well as better confidence calibration, which is critical for medical-related tasks.
Therefore, unifying multiple speech elicitation paradigms within a single model not only improves detection performance and modelling efficiency, but also facilitates more practical and scalable deployment of speech-based suicide risk detection systems.

\section{Limitations and Ethical Considerations}

Our conclusions are derived from the MINI-KID scale that gauges current suicide risk based on respondents' immediate answers. Though MINI-KID is widely used and often treated as a benchmark for adolescent assessment, it cannot fully capture the multifaceted nature of suicide risk. Accordingly, these findings should not be interpreted as forecasts of future suicidal behaviour and apply only within the context of this evaluation.

\newpage

\section{Acknowledgments}
This work was supported by the National Natural Science Foundation of China (Grant No. 62501336).

\section{Generative AI Use Disclosure}
Generative AI tools were used for language polishing and editing of the manuscript. All authors have reviewed and approved the final content, ensuring accuracy and accountability for the work.

\bibliographystyle{IEEEtran}
\bibliography{refs}

\end{document}